\documentclass{iopart}
\usepackage{iopams}
\usepackage{graphicx}
\begin{document}

\title{
Magnetic properties of a $S=1/2$ zigzag spin chain compound ${\rm (N_2H_5)CuCl_3}$
}

\author{N. Maeshima\dag\footnote[5]{Present address: Large-scale Computational Science Division, Cybermedia Center, Osaka University 1-32 Machikaneyamacho, Toyonaka Osaka 560-0043, Japan}
 , M. Hagiwara\ddag, Y. Narumi\S, K. Kindo\S,  T. C. Kobayashi\S\ and K. Okunishi$\|$}

\address{\dag\ Faculty of Engineering, Osaka Electro-Communication University, 
Neyagawa, Osaka 572-8530, Japan}

\address{\ddag\ RIKEN(The Institute of Physical and Chemical Research), Wako, Saitama 
351-0198, Japan}

\address{\S\ KYOKUGEN, Osaka University, Toyonaka, 
Osaka 560-8531, Japan }

\address{$\|$\ Department of Physics, Faculty of Science, Niigata University, Niigata 
950-2181, Japan }


\begin{abstract}
We present a theoretical and experimental study of a quasi-one-dimensional zigzag antiferromagnet ${\rm (N_2H_5)CuCl_3}$, which can be viewed as weakly coupled Heisenberg chains with a frustrated interaction. We first discuss generic features of the magnetic properties of the zigzag spin chain between the nearly single chain case and the nearly double chain case, on the basis of the finite temperature density-matrix renormalization group (DMRG) calculations.  We next show the experimental results for the magnetic susceptibility and the high-field magnetization of a single crystal of ${\rm (N_2H_5)CuCl_3}$ above the N\'eel temperature $T_{\rm N}=1.55{\rm K}$. By comparing the experimental data with the DMRG results carefully, we finally obtain the ratio of the nearest and next-nearest exchange couplings as $J_1/J_2=0.25$ with $J_2/k_{\rm B}=16.3{\rm K}$. We also investigate the three-dimensional (3D) coupling $J'$ effect by using the mean-field theory combined with the DMRG calculations. The estimated value $J' \sim 0.04J_2$ supports good one-dimensionality of ${\rm (N_2H_5)CuCl_3}$ above $T_{\rm N}$.
\end{abstract}


\submitto{\JPCM}

\maketitle

\section{Introduction}

\label{sec:intro}
Frustration effects in low-dimensional antiferromagnetic (AF) quantum spin
systems have attracted much attention because of their peculiar behavior in 
low-energy physics. The quantum effect and the geometrical frustration induce 
strong fluctuation cooperatively, which often gives rise to various 
non-magnetic ground states and quantum phase transitions~\cite{MG,haldane,delta,gelfand,lhuillier}.

\begin{figure}
\begin{center}
\includegraphics[width=7.5cm,clip]{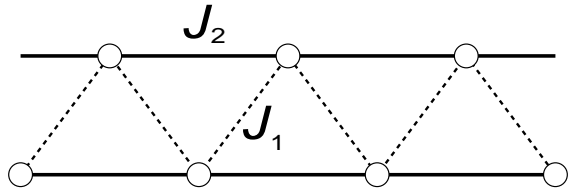}
\end{center}
\caption{The zigzag spin chain. The dashed lines and solid lines denote the 
nearest neighbor coupling $J_1$ and the next-nearest neighbor coupling 
$J_2$ respectively.} 
\label{fig:chain}
\end{figure}

A typical example of such frustrated low-dimensional magnets is the 
$S=1/2$ antiferromagnetic zigzag chain (see figure~\ref{fig:chain}). The 
Hamiltonian of the zigzag spin chain is given by
\begin{equation}
{\cal H}= \sum_i \left[ J_1{\vec S}_i\cdot{\vec S}_{i+1} + J_2{\vec S}_i\cdot {\vec 
S}_{i+2}\right] - g\mu_{B}H\sum_iS^z_i,
\end{equation}
where ${\vec S}$ is the $S$=$1/2$ spin operator, $g$ is the Land\'{e}'s $g$ factor and $\mu_B$ is the Bohr magneton.
$J_1$ and $J_2$ denote the nearest neighbor and the next-nearest neighbor 
coupling constants respectively, and $H$ is the applied magnetic field. 
Intensive theoretical studies have shown that the zigzag chain has a rich phase  structure at the zero magnetic field. 
When $J_2=0$, the system is equivalent to the Heisenberg chain having the gapless excitation~\cite{dCP}.
Increasing $J_2$, a quantum phase transition from the critical spin liquid 
phase to the gapped dimer phase~\cite{haldane,tone} occurs at $J_2/J_1=0.2411$~\cite{nomura}.
This gapped dimer phase is expected to extend up to the limit  $J_2/J_1=\infty$~\cite{whiteaffleck}, which corresponds to the two decoupled Heisenberg chains.
Here, it should be remarked that  the dimer gap $\Delta$ for  $J_1 \ll J_2$ 
is exponentially small: $\Delta\sim \exp(- {\rm const} \times
J_2/J_1)$~\cite{whiteaffleck}.

The theoretical studies mentioned above have stimulated experimental  studies of the zigzag spin chain. In fact, the magnetic properties of ${\rm SrCuO_2}$~\cite{matsuda} and Cu[2-(2-aminomethyl)pyridine]~\cite{kikuchi} were investigated recently. 
For the latter case,  the nearly gapless behavior of the magnetic susceptibility was observed for a powder sample, and then the ratio of the couplings of Cu[2-(2-aminomethyl)pyridine] is estimated as $J_2/J_1 \sim 0.2$~\cite{kikuchi}.
In this paper, we focus on  another type of the zigzag compound  $\rm (N_2H_5)CuCl_3$~\cite{brown}, which is interestingly expected to have the coupling constant  $J_1 \ll J_2$~\cite{hagiwara}.  Since  $\rm (N_2H_5)CuCl_3$ can be synthesized as a single crystal,  we can  precisely compare its magnetic properties  with numerical calculations.

An important point on the compound is that its magnetic property is very similar to the one in the single Heisenberg chain limit, although it has the double chain structure. Thus how we can distinguish the magnetic behavior of the zigzag chains between $J_1 \gg J_2$ and $J_1 \ll J_2$  is an important problem  from both theoretical and experimental views. In what follows, we study the magnetic properties of a single crystal sample of $\rm (N_2H_5)CuCl_3$, on the basis of the precise comparison between the numerical results obtained by the finite temperature density matrix renormalization group (DMRG) method~\cite{ftdmrg1,ftdmrg2} and the experimental measurements of the magnetic susceptibility and the high-field magnetization.
In addition, we remark the effect of the three-dimensional (3D) coupling($J'$)~\cite{3d} on $\rm (N_2H_5)CuCl_3$; we apply the mean-field theory combined with the finite temperature DMRG to verify the one dimensionality of $\rm (N_2H_5)CuCl_3$~\cite{klumper,nishiyama}.

In section~\ref{sec:exp}, we describe the finite temperature DMRG method and the fundamental aspects of the single crystal of $\rm (N_2H_5)CuCl_3$. In section~\ref{sec:dmrg}, we show the DMRG results for the susceptibility and the magnetization curve, and discuss generic features of them. In section~\ref{sec:result} we show the experimental data of the magnetic susceptibility and the high-field magnetization parallel to the chain direction. Then the exchange coupling $J_2/k_{\rm B}=16.3$K and the ratio $J_1/J_2=0.25$ are obtained by the precise comparison of the experimental data with the DMRG results. We also discuss the 3D effect and estimate the coupling $J'\sim 0.04J_2$. In section~\ref{sec:dis} the conclusion is summarized.


\section{Computational and experimental details}
\label{sec:exp}
\subsection{finite temperature DMRG}

For the reliable analysis of the experimental results, systematic numerical data of the magnetization curve and the susceptibility are required in a wide parameter region. However, the quantitative calculation for the zigzag chain at finite temperatures has not been made yet especially for the nearly two chain cases.
In this study  we employ the finite temperature DMRG method~\cite{ftdmrg1,ftdmrg2}, which is free from the negative sign problem and thus enables us to calculate the quantities sufficiently down to low temperatures with high accuracy. 
We note that the method is actually applied to some frustrated spin ladder systems successfully~\cite{klumper,maisinger,maeoku,kagome}.
Following  the implementation procedure described in reference~\cite{maeoku}, we calculate the magnetization and the susceptibility with the maximum number of the retained bases $m=240$. The lowest temperature we can reach is $k_{\rm B}T/J=0.07$, where $J$ is the larger one of the two exchange couplings $J_1$ and $J_2$.
We have confirmed that the computed data converge with respect to $m$ and the Trotter number.

\subsection{experiment}

\begin{figure}
\begin{center}
\includegraphics[width=8cm,clip]{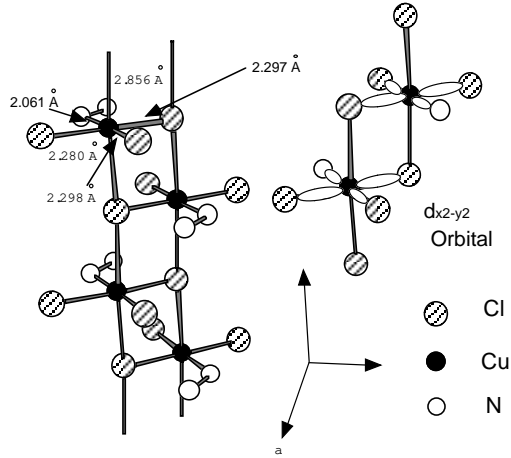}
\end{center}
\caption{Chain structure of $\rm (N_2H_5)CuCl_3$.} 
\label{fig:crystal}
\end{figure}

We next summarize the synthesis and the crystal structure of $\rm (N_2H_5)CuCl_3$.
$\rm (N_2H_5)CuCl_3$ samples were synthesized according to the method described
in reference~\cite{brown}.  After several syntheses, we obtained small single crystal samples 
with a typical size of $1.5\times 2\times 3 mm^3$.
Chemical analysis shows good agreement between the observed and calculated 
ratios of the elements. This compound crystallizes in the orthorhombic 
system (space Group: $\it{P}$nma)~\cite{brown}. The lattice constants at room 
temperature are $a=14.439(2)$\AA, $b=5.705(1)$\AA, and $c=6.859(1)$\AA. 
The chain structure of this compound is shown in figure~\ref{fig:crystal}.
This compound has the ladder structure in which Cu and Cl align 
alternately along the leg, forming a copper zigzag chain. 
Copper 3d hole orbitals (3d$_{x^2-y^2}$) are situated perpendicularly to the leg, resulting in small exchange interactions along the leg.

Magnetic susceptibilities were measured with a SQUID magnetometer (Quantum Design's MPMS-XL7) installed at KYOKUGEN in Osaka University. High-field magnetization measurements up to 30T were carried out with a pulse magnet at the same place. Besides, X-band ESR (Varian E109 spectrometer) and specific heat measurements were carried out at the same place.

\section{DMRG results}
\label{sec:dmrg}

In this section, we present the DMRG results for the susceptibility and the magnetization curve, and then discuss the frustration dependence of such quantities for the nearly single chain and nearly decoupled chains cases.

\begin{figure}[ht]
\begin{center}
\includegraphics[width=8cm,clip]{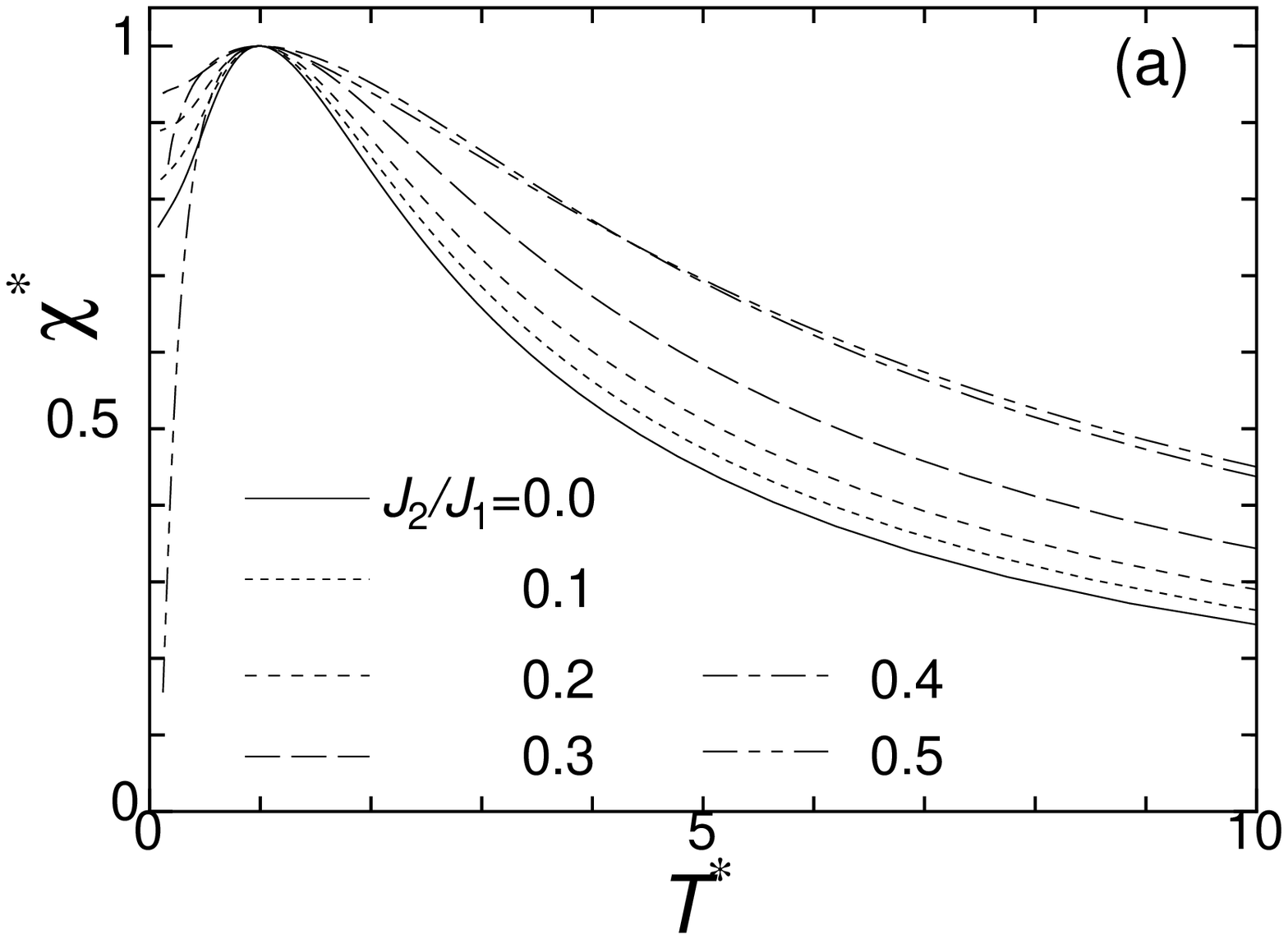}
\includegraphics[width=8cm,clip]{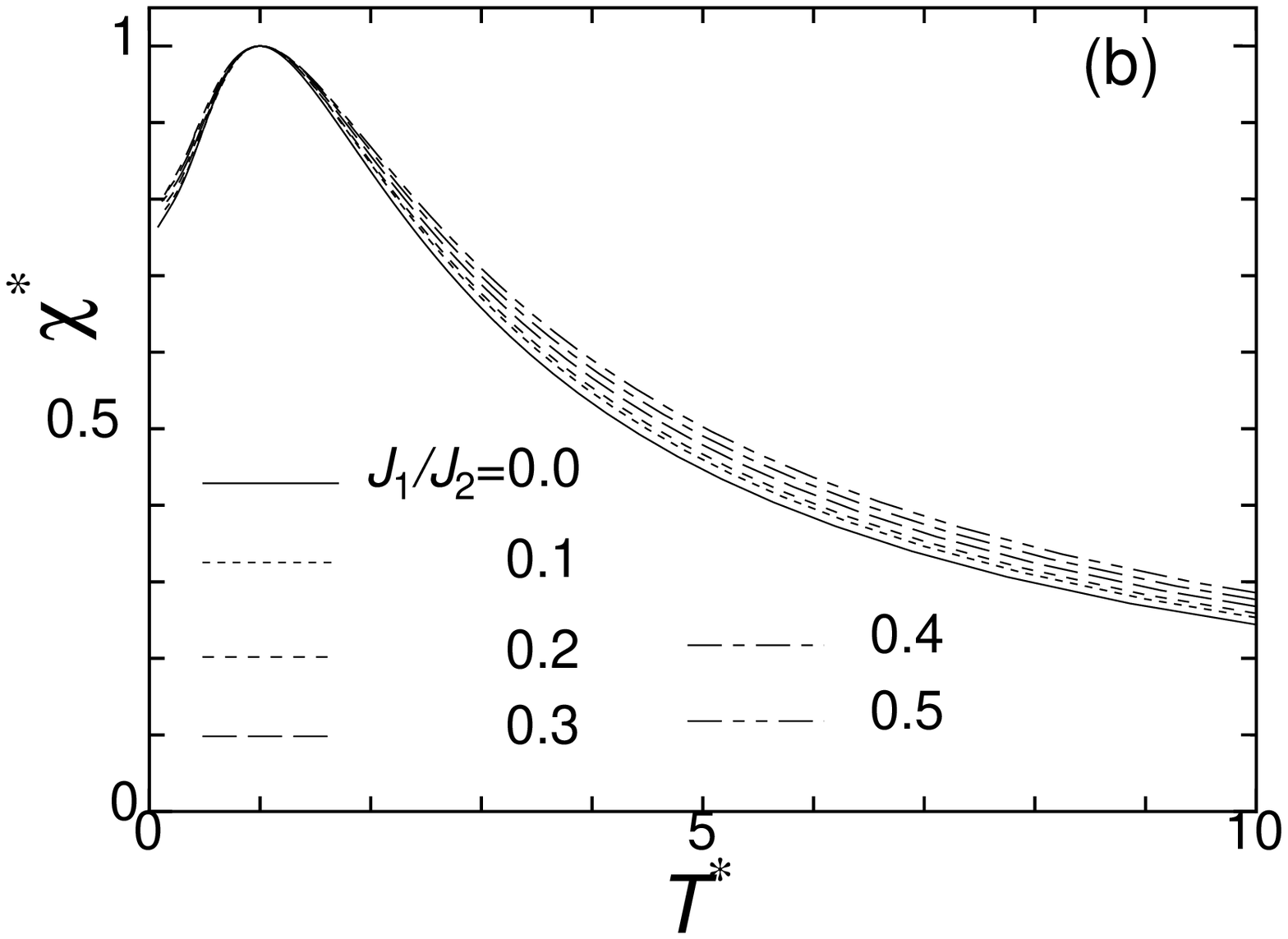}
\end{center}
\caption{Finite temperature DMRG results of the susceptibility (a) for $J_1>J_2$ and (b) for $J_1<J_2$.}
\label{fig:chid}
\end{figure}

 Figure~\ref{fig:chid} shows the normalized susceptibilities $\chi^*\equiv \chi/\chi_{\rm max}$, where $\chi$ is the magnetic susceptibility per site and $\chi_{\rm max}$ is its maximum value. 
The normalized temperature $T^*\equiv T/T_{\rm max}$ is also introduced, where $T_{\rm max}$ is defined as the temperature at which $\chi$ indicates $\chi_{\rm max}$.  The values of $T_{\rm max}$ and $\chi_{\rm max}$ are summarized in table 1.

\begin{table}[bh]
\caption{The values of $T_{\rm max}$ and $\chi_{\rm max}$ extracted from the DMRG results with the normalization $J_1=1$ for $J_1>J_2$ and $J_2=1$ for $J_1<J_2$. We also set $k_{\rm B}=g\mu_B=1$ for simplicity.}
\begin{indented}
\item[]
\begin{tabular}{@{}lllllll} 
\br
\multicolumn{3}{c}{ $J_1>J_2$ }&& \multicolumn{3}{c}{ $J_1<J_2$ } \\ 
\mr
    $J_2/J_1$ & $T_{\rm max}$ & $\chi_{\rm max}$ && $J_1/J_2$ &$T_{\rm max}$ & $\chi_{\rm max}$ \\
\mr
       0.0  &  0.640  &   0.147    &&   0.0   & 0.640  & 0.147  \\
       0.1  &  0.588  &   0.147    &&   0.1   & 0.640  & 0.143  \\
       0.2  &  0.516  &   0.148    &&   0.2   & 0.635  & 0.139  \\
       0.3  &  0.419  &   0.149    &&   0.3   & 0.630  & 0.135  \\
       0.4  &  0.299  &   0.152    &&   0.4   & 0.620  & 0.131  \\
       0.5  &  0.279  &   0.153    &&   0.5   & 0.606  & 0.128  \\ 
\br
\end{tabular}
\end{indented}
\end{table}

Let us first consider the nearly single chain case ($J_1>J_2$). In figure 3(a), we can see the feature of the gapless Heisenberg chain for $J_2/J_1 \le 0.3$; in a high temperature region $\chi^*$ increases with decreasing $T^*$, and takes a round peak at $T^*=1$. Then $\chi^*$ starts to decrease, and finally converges to a finite value as $T^* \to 0$.
Although the spin fluid-dimer transition occurs at $J_2/J_1=0.2411$, the expected dimer gap is exponentially small for $J_2/J_1 = 0.3$~\cite{nomura}, where the almost gapless behavior of $\chi^*$ appears~\cite{maeoku}.
As $J_2/J_1$ is increased to 0.4 and 0.5, the dimer-gapped phase starts to develop rapidly, and then the exponential decay of $\chi^*$ appears explicitly in the low temperature region. 
This rapid growth of the dimer gap also causes the considerable enhancement of $\chi^*$ in the high temperature side ($T^*>1$) for $J_2/J_1=$0.4 and 0.5.

We next consider the nearly decoupled chains case~($J_1<J_2$), where we can also see that $\chi^*$ basically shows the feature of the gapless chain.
A remarkable point in contrast to the nearly single chain case is that the gapless-like behavior is observed in the wide range of $J_1/J_2$ (see figure 3(b)) and thus the dependence of $\chi^*$ on $J_1/J_2$ is relatively weak.
This wide-range gapless-like behavior is attributed to the extremely small spin gap: $\Delta \sim \exp( - {\rm const}\times J_2/J_1)$, although the system is considered to be the dimer gapped phase for $J_1\ne 0$.
For example, the spin gap is estimated as $0.03J_2$ even at $J_1/J_2=0.5$~\cite{whiteaffleck}, and thus the effect of the gap is easily smeared out by the finite temperature effect.

Here, we wish to point out an important feature of the susceptibility when analyzing
the experimental results. If the frustration is small, it may be difficult to 
discriminate between the behavior of the nearly single chain and that of the 
nearly decoupled chain  only from the susceptibility result, because $\chi^*$ of the nearly single chain and that of the nearly decoupled chain are very similar to each other.
However, an important point is that the frustration dependence of  $T_{\rm max}$ for $J_1>J_2$ is relatively sensitive to $J_2/J_1$(see table 1), implying that the energy scale of the dominant coupling constants has a certain difference between $J_1 > J_2$ and $J_1 <J_2$, even if the shapes of the susceptibility are quite similar to each other. 
 For example, the shape of $\chi^*$ for $J_2/J_1=0.1$ is almost coincident with that  for $J_1/J_2=0.3$, but the values of $T_{\rm max}$  for  $J_2/J_1=0.1$ and $J_1/J_2= 0.3$ can read as 0.588 and 0.630 respectively, which show about 7\% deviation. 

We now proceed to the results of the magnetization curves.
Figure 4 shows the magnetization curves both for $J_1>J_2$ and for $J_1<J_2$ at $k_{\rm B}T/J_1$($J_2$)=0.125, which can be directly compared with experimental results of
(N$_2$H$_5$)CuCl$_3$ measured at 2K as shown later.
In the figure, we can basically see the characteristic property in the gapless Heisenberg chain as well;
after the linear behavior  with $H$ is observed near the zero magnetic 
field ($H=0$), the magnetization $M$ increases rapidly in the high field region, and finally saturates near $g\mu_BH/J_1(J_2)\sim 2.5$. 
However, what we want to remark here is that the quantitative differences appear in the magnetization curve. 
For instance, in figure 4(a), we can see that the growth of the dimer gap for $J_2/J_1 \ge 0.4$ causes the enhancement of the magnetization in the middle-field region ($g\mu_BH/J_1 \simeq 1$), which is consistent with the susceptibility result.
Moreover the magnetization curve around the saturation field $H_s$ is shifting to the high field side systematically as the frustration increases.

\begin{figure}[hb]
\begin{center}
\includegraphics[width=7.8cm,clip]{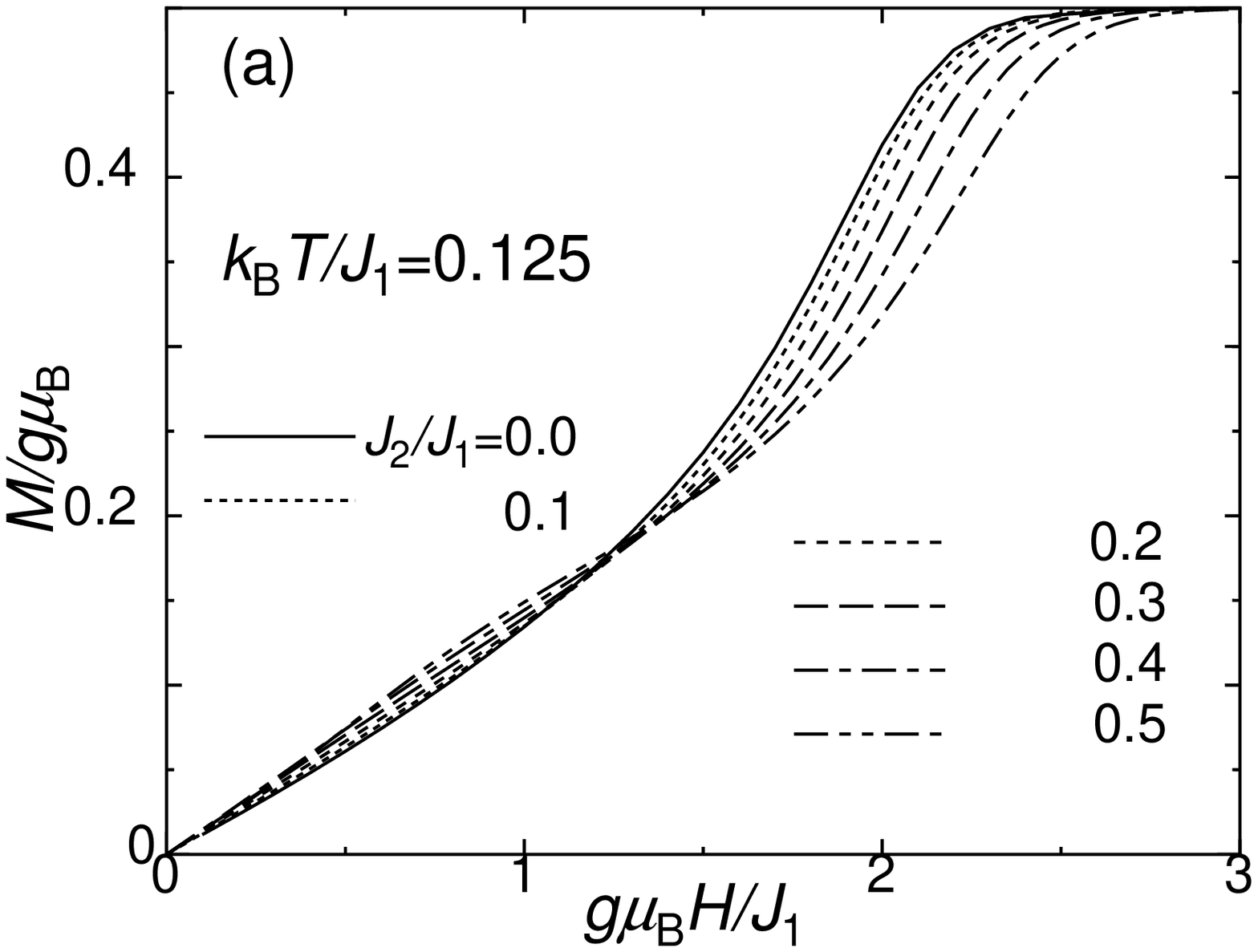}
\includegraphics[width=7.8cm,clip]{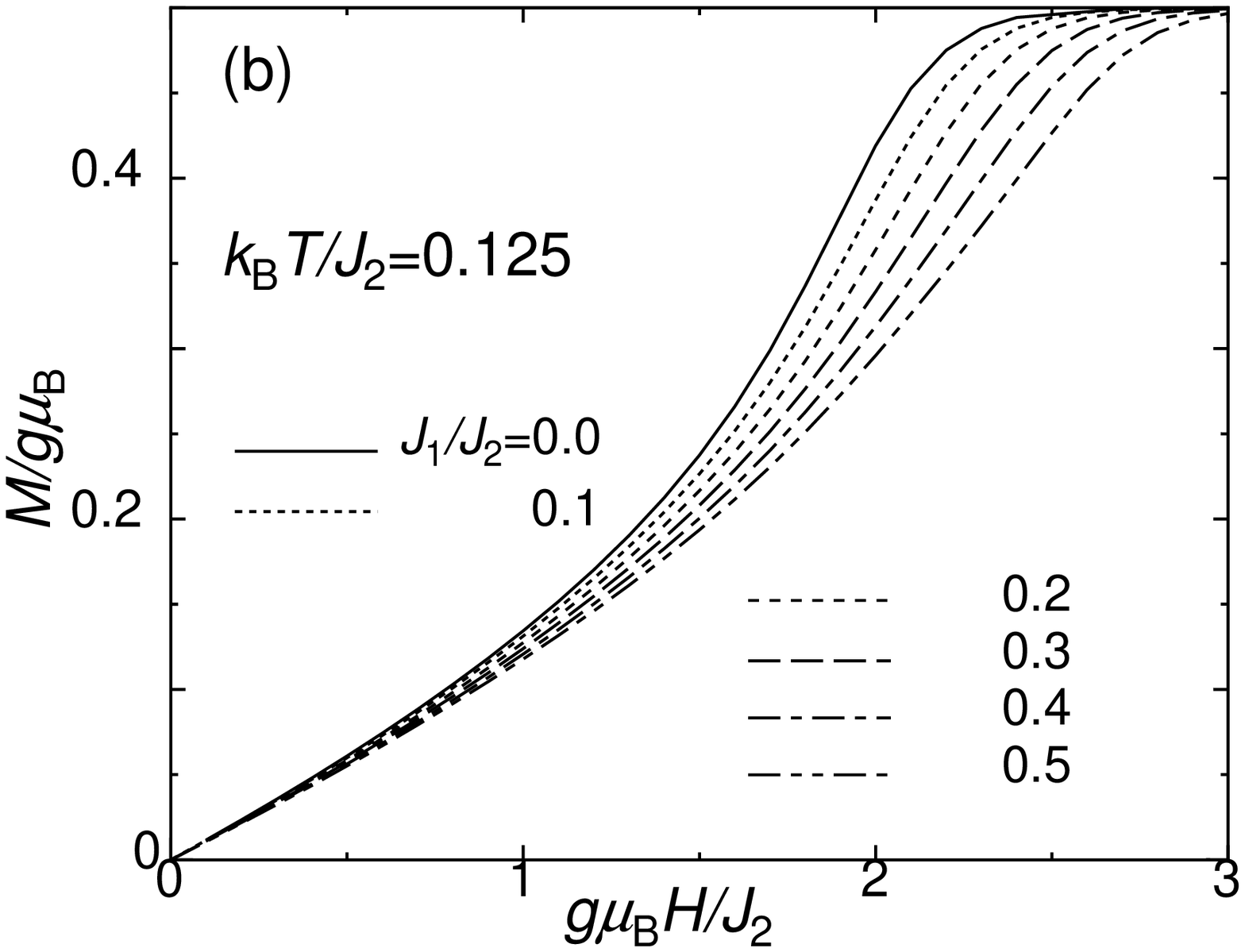}
\end{center}
\caption{The magnetization curves obtained with the DMRG (a) for $J_1>J_2$ at $k_{\rm B}T/J_1=0.125$ and (b) for $J_1<J_2$ at $k_{\rm B}T/J_2=0.125$.} 
\label{fig:mhd}
\end{figure}

In order to capture the feature of the magnetization curves in each region of $J_1 > J_2$ and $J_1<J_2$, we here focus on the above mentioned systematic shift of the magnetization curve near the saturation field $H_s$.
For the zigzag chain, the behavior of the saturation field $H_s$ is easily obtained as
\begin{equation}
H_s = 
\cases{
\frac{2J_1}{g\mu_B} & for $( J_2/J_1<1/4)$  \\
 \frac {1}{g\mu_B}( J_1+2J_2+\frac{J_1^2}{8J_2}) & for $( J_2/J_1>1/4)$ . 
}
\end{equation}
For the nearly single chain, the saturation field $H_s$ stays at $2J_1/g\mu_{\rm B}$ for $J_2/J_1<1/4$, and it starts to increase gradually as $J_2/J_1$ increases beyond $1/4$ . 
On the other hand, for the nearly decoupled chain, $H_s$ starts to increase linearly  as soon as the inter-chain coupling is introduced.
This dependence of $H_s$ readily appears in the magnetization curve shown in figure~\ref{fig:mhd}; 
the magnetization curves near $H_s$ for  $J_2/J_1<0.3$ remains at the ``vicinity'' of the pure Heisenberg one. 
In contrast, for the nearly double chain, we can see that the curves near $H_s$ shift to the high field side linearly with respect to $J_1/J_2$. 
As a result, the magnetization curves near $H_s$ for $J_1>J_2$ become slightly steeper than those for $J_1<J_2$.

As was seen for the susceptibility result,  the dominant coupling  determined from the susceptibility exhibits a slightly different value depending on the nearly chain or  nearly decoupled chains cases.
Then such difference of the dominant coupling affects the magnetization curve near the saturation field, which can be resolvable with the comparison between  the accurate DMRG result and the experimental ones.
Therefore  the coupling constant of the zigzag chain can be expected to be determined with the cooperative use of the susceptibility and the magnetization curve.

\section{Experimental results}
\label{sec:result}

\subsection{ESR and specific heat}

We performed X-band ESR measurements at room temperature on a single 
crystal sample of (N$_2$H$_5$)CuCl$_3$ in order to get the $g$-values 
precisely.   Accurate comparison 
between the experimental data and the numerically calculated ones 
requires the precise $g$-values.  Figure 5 shows the angular dependence 
of the $g$-value.  We obtained the $g$-values of 2.285 and 2.060 for 
the external magnetic field parallel and perpendicular to the chain, respectively. In the 
following comparisons between experiments and calculations, we used 
these $g$-values.

We carried out specific heat measurements on a powder sample of 
(N$_2$H$_5$)CuCl$_3$.  Figure~6 shows the temperature dependence of 
the specific heat divided by temperature. Here, the data include the 
lattice contribution.  We observed an anomaly at 1.55K ($T_{\rm N}$) which 
is probably due to the antiferromagnetic long-range ordering. In order 
to avoid the 3D effect in this quasi-one-dimensional system, high field magnetization experiments were done at 2K above $T_{\rm N}$.
\begin{figure}
\begin{center}
\includegraphics[width=8cm,clip]{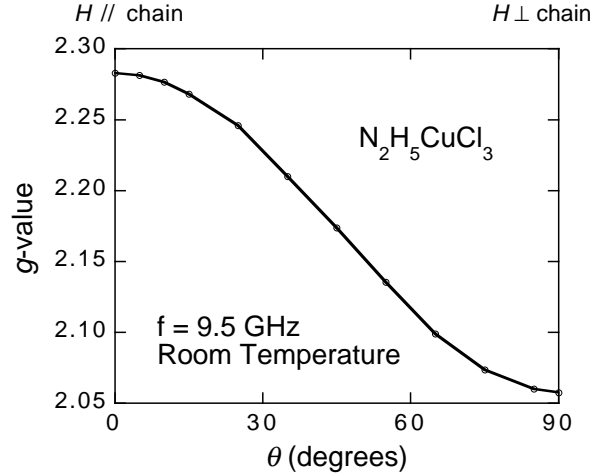}
\end{center}
\caption{The angular dependence of the $g$-value of  $\rm (N_2H_5)CuCl_3$.} 
\label{fig:esr}
\end{figure}

\subsection{magnetic susceptibility}

\begin{figure}
\begin{center}
\includegraphics[width=8cm,clip]{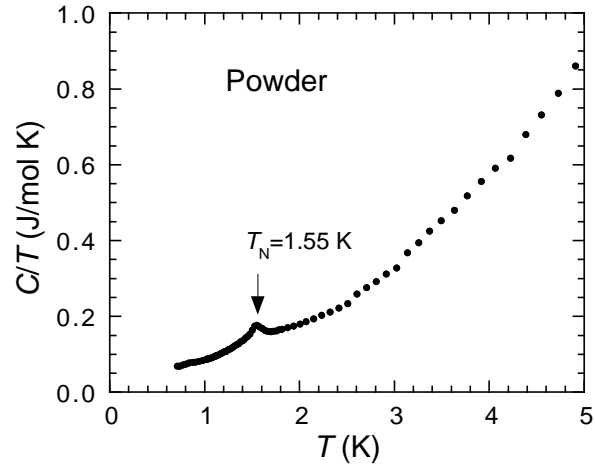}
\end{center}
\caption{The specific heat of a powder sample of $\rm (N_2H_5)CuCl_3$.} 
\label{fig:heat}
\end{figure}

\begin{figure}
\begin{center}
\includegraphics[width=8cm,clip]{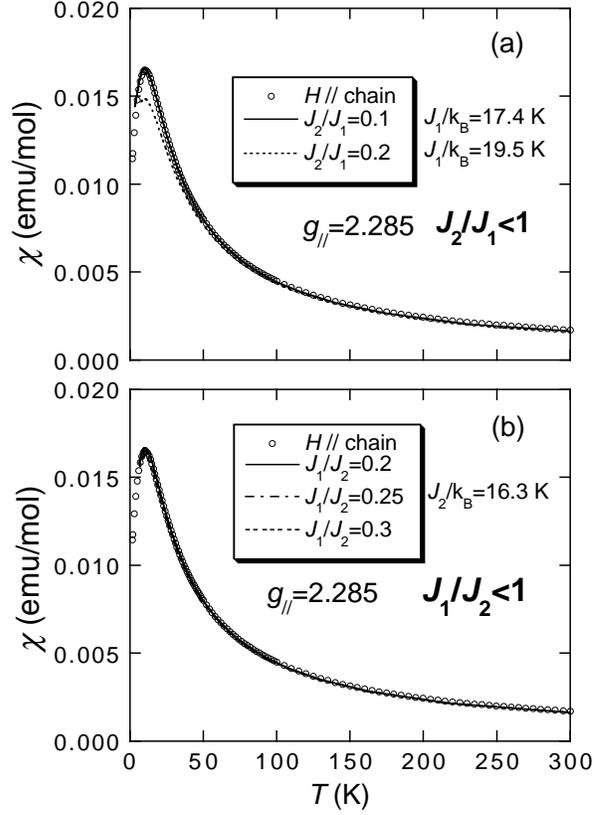}
\end{center}
\caption{Magnetic susceptibility of $\rm (N_2H_5)CuCl_3$. The open circles denote experimental data $\chi_{\scriptscriptstyle \rm EXP}$ and the lines show finite temperature DMRG results $\chi_{\scriptscriptstyle \rm DMRG}$.}
\label{fig:sus}
\end{figure}

Magnetic susceptibility $\chi_{\scriptscriptstyle \rm  
EXP}$ of the single crystal of $\rm (N_2H_5)CuCl_3$ parallel to the 
chain, which was used for the ESR measurement, is presented  in 
figure~\ref{fig:sus}, where the DMRG results $\chi_{\scriptscriptstyle \rm 
DMRG}$ are also shown for comparison.
In the high temperature region the $\chi_{\scriptscriptstyle \rm EXP}$ 
exhibits the Curie-Weiss-type $T$-dependence, and has a round peak around 
$T=10{\rm K}$.
When the temperature is decreased further, the $\chi_{\scriptscriptstyle \rm EXP}$ also 
decreases and finally converges to a finite value in the zero temperature 
limit.
Since these features are characteristic  in gapless 1D antiferromagnets, 
we should take account of the two possibilities of the gapless zigzag chains, as mentioned in the introduction.
We thus analyze $\chi_{\scriptscriptstyle \rm  EXP}$ with the DMRG results 
both for the single chain limit  ($J_1 > J_2$) and the weakly coupled 
Heisenberg chain limit ($J_1 < J_2$). 
In fitting, we first estimate the dominant exchange coupling constant from the 
position of the broad peak, by using $g$=2.285 obtained from the ESR, and 
then tune another exchange coupling constant so as to reproduce the shape of the susceptibility well.

For the former case ($J_1>J_2$), we find that $\chi_{\scriptscriptstyle 
\rm  DMRG}$ of $J_2/J_1 =0.1$ with $J_1/k_{\rm B}=$17.4K agrees well with 
$\chi_{\scriptscriptstyle \rm  EXP}$ in the whole temperature region (see 
figure~\ref{fig:sus} (a)). 
For the latter ($J_1<J_2$),  we can also see  that the DMRG results 
for $ 0.2 <J_1/J_2 <0.3$ well reproduce the shape of the 
experimental susceptibility  as in figure~\ref{fig:sus} (b), where the  best 
fitted value is obtained as $J_1/J_2=0.25$ with $J_2/k_B=16.3$K.
Since the DMRG results seem to explain the whole shape of the 
susceptibility well for both cases, it is  unfortunately difficult to confirm which case is realized in $\rm (N_2H_5)CuCl_3$ only from the susceptibility result, as discussed in the previous section.
However, we note that the dominant coupling $J_1/k_B=$17.4K for $J_1>J_2$ is about 6\% bigger than $J_2/k_B=16.3$K for $J_1<J_2$, which is consistent with the DMRG result. 
For the case of $H$ $\perp$ chain, very similar results were obtained using the same  parameter values of the exchange couplings as for $H$ // chain and $g$=2.060 from ESR. 

Before proceedings to the next subsection we make a comment about a
possibility of the weak ferromagnetic coupling. We have also computed the 
case of a negative $J_2$($J_1$) for $J_1>J_2$($J_1<J_2$). 
However, the shape of the susceptibility is changed drastically  by 
introducing the ferromagnetic coupling, and so that the couplings of $\rm 
(N_2H_5)CuCl_3$ are antiferromagnetic. 

\subsection{magnetization curve}

\begin{figure}[th]
\begin{center}
\includegraphics[width=8cm,clip]{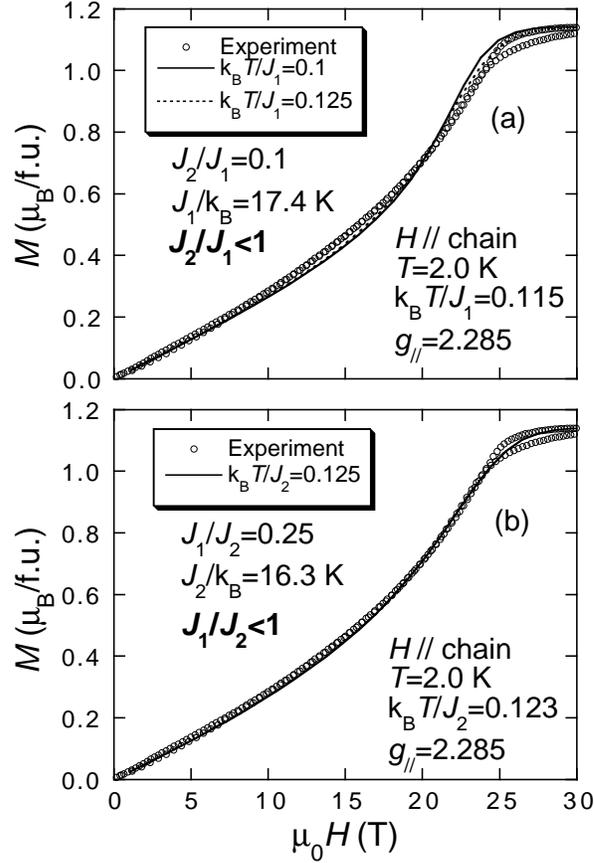}
\end{center}
\caption{The magnetization curves of $\rm (N_2H_5)CuCl_3$. The
comparison with the DMRG results (a) for $J_2/J_1=0.1$ and (b) for
$J_1/J_2=0.25$.}
\label{fig:mag}
\end{figure}

Figure~\ref{fig:mag} shows the magnetization curve for the single crystal of 
$\rm (N_2H_5)CuCl_3$ along the chain at $T=2.0$K up to 30T. The temperature was 
set a little above $T_{\rm N}$ (1.55K) as already mentioned. Around the saturation field,
there is a small hysteresis of magnetization between the ascending and 
descending processes, due to the magnetocaloric effect. In the figure, we can see that the measured curve of  $\rm (N_2H_5)CuCl_3$ is qualitatively similar to the gapless chain, which is consistent with 
the susceptibility results. Using the parameters estimated in the previous subsection, we now discuss 
the magnetization processes of the DMRG and the experiments. The temperature used in the numerical calculations was determined close to that in the experiments.

The results of comparisons are also shown in figure~\ref{fig:mag}, where the DMRG curves both for $J_1>J_2$ and for $J_1<J_2$ are in good agreement with the  experimental one in the low-field region. However, the magnetization curve in the high field region is reproduced well by the coupling of $J_1/J_2=0.25$, as can be seen in figure~\ref{fig:mag}(b). Indeed the gradient of the calculated curve for $J_1>J_2$ is slightly bigger than the experimental one around $15< H <25$T, while the curve for $J_1<J_2$ is well fitted in the all range of the magnetic field. The difference between two cases described in the previous section certainly appears in high-field region. Hence we have determined that the zigzag chain compound $\rm (N_2H_5)CuCl_3$ is a weakly coupled double chain system with the parameters $J_1/J_2=0.25$ and $J_2/k_{\rm B}=16.3$K.

\subsection{3D interaction}
\label{sec:mft}

\begin{figure}[ht]
\begin{center}
\includegraphics[width=8cm,clip]{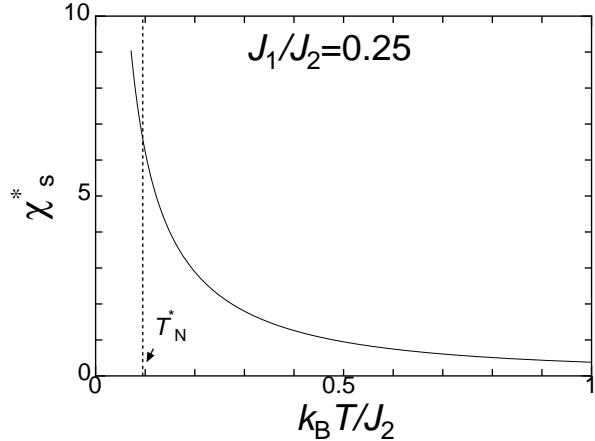}
\end{center}
\caption{Staggered susceptibility of the zigzag chain for $J_1/J_2=0.25$. The dotted line denotes the normalized transition temperature $T^*_N \equiv k_{\rm B}T_{\rm N}/J_2=0.0951\cdots$ for $\rm (N_2H_5)CuCl_3$. The value of $\chi^*_s$ at $k_{\rm B}T/J_2=T^*_{\rm N}$ gives the coupling $zJ'$ following equation~(\ref{eq:mf}).}
\label{fig:chis}
\end{figure}

In this subsection, we would like to discuss the effect of the 3D interaction on $\rm (N_2H_5)CuCl_3$.
Since the specific heat shows the 3D ordering at $T_{\rm N}=1.55$K, some consideration for the 3D coupling may be required.
To this end, we employ here the mean-field theory combined with the DMRG~\cite{klumper,nishiyama}.
We use the obtained values of the couplings $J_1/J_2=0.25$ and $J_2/k_{\rm B}=16.3$K, and assume a 3D Heisenberg type interaction whose coupling constant is denoted as $J'$.
According to the strong leg interaction $J_2$, we assume the staggered magnetic field along the leg direction, and then determine the assumed magnetic field and the DMRG-calculated magnetic order self-consistently.
In particular, the phase transition point within the mean field theory is given by
\begin{equation}
\chi_s(T_{\rm N}) = \frac{1}{zJ'}\label{eq:mf}
\end{equation}
where $\chi_s(T)$ is the staggered susceptibility and $z$ is the coordination number.
Calculating $\chi_s$ with the finite temperature DMRG, we can estimate $J'$ from the experimentally determined N\'{e}el temperature $T_{\rm N}$ through the equation~(\ref{eq:mf}). 
In figure~\ref{fig:chis}, we plot the normalized staggered susceptibility $\chi^*_s \equiv J_2\chi_s$ obtained with the finite temperature DMRG.
In the figure, we find that the crossing point of $\chi^*_s$ and the experimentally obtained N\'{e}el temperature is given by $\chi^*_s=6.6$ and $T^*_N \equiv k_{\rm B}T_{\rm N}/J_2 \sim 0.095$, where the $T_{\rm N}^*$ is the normalized transition temperature. Assuming $z$=4, we obtain $J'\sim 0.04J_2$($\sim 0.6$K), which is about $0.2J_1$. Thus we can conclude that the 3D interaction effect is much smaller than the intra-chain couplings. Therefore the zigzag chain picture is valid for  $\rm (N_2H_5)CuCl_3$ above the N\'{e}el temperature.

\section{Summary and discussion} 
\label{sec:dis}

We have investigated magnetic properties of $\rm (N_2H_5)CuCl_3$, which can be regarded as a zigzag chain with the coupling $J_1<J_2$ by the accurate comparison between the numerical calculations and experiments.

First, we have calculated the susceptibility and the magnetization curve of the zigzag chain using the finite temperature DMRG method for the nearly single chain case $J_1>J_2$ and for the nearly double chain case $J_1<J_2$  to clarify the differences between the two cases appearing in the parameter dependence of the physical quantities. From the experimental point of view, the quantitative difference appearing in the high-field region of the magnetization curve is found to help us to identify the magnetic nature of $\rm (N_2H_5)CuCl_3$.

We have next presented the experimental results for the single crystal of $\rm (N_2H_5)CuCl_3$. The $g$-values parallel and perpendicular to the chain are determined as $g=2.285$ and 2.060, respectively by the ESR experiment.  The N\'{e}el temperature is also estimated as $T_{\rm N}=1.55$K from the specific heat of a powder sample. Above the N\'{e}el temperature, the observed susceptibility and the magnetization for the single crystal show gapless or nearly gapless behavior. An accurate comparison of the experimental data with the DMRG results shows that the DMRG results for $J_1/J_2=0.25$ with $J_2/k_{\rm B}=16.3$K well reproduce the experimental results. The estimated couplings also show that the dimer gap of $\rm (N_2H_5)CuCl_3$ is $0.001 J_2 \sim 0.02$K~\cite{whiteaffleck}, which is much smaller than $T_{\rm N}$. Hence the gapless behavior of $\rm (N_2H_5)CuCl_3$ observed above $T_{\rm N}$ is consistent with the obtained exchange couplings.

Finally, we have discussed the effect of the 3D interaction, using the mean-field theory combined with the finite temperature DMRG. The estimated value of the 3D coupling $J'$ ($\sim 0.04J_2$) is sufficiently small compared with the intra-chain couplings, which supports the zigzag chain picture of the compound.

In the present paper, we have addressed the zigzag compound of $\rm (N_2H_5)CuCl_3$ in terms of the bulk quantities: the magnetic susceptibility and the high-field magnetization, for which the rather precise analyses are required. However, the frustration effect often gives rise to the incommensurate behavior in the correlation function~\cite{whiteaffleck,incomme}, which is difficult to be captured by the bulk quantities. Thus it is also an interesting problem to investigate the compound by measurements of the fluctuation-associated quantities with the position resolvable methods, such as NMR or the inelastic neutron scattering.


\ack

This work was partly supported by the Grand-in-Aid for Scientific 
Research from the Japanese
Ministry of Education, Culture, Sports, Science and Technology. The experimental work 
was carried out under visiting Researcher's Program of KYOKUGEN, Osaka 
University.



\begin{thebibliography}{99}
\bibitem{3d} The 3D coupling noted here is usually called  ``inter-chain coupling''. However this naming is rather confused with ``inter-chain coupling'' originating from the double chain structure of the zigzag interaction.  In this paper, we thus call the 3D inter-chain coupling ``3D coupling''.
\end{thebibliography}
\end{document}